\documentclass[prl,twocolumn,showpacs,preprintnumbers,amsmath,amssymb]{revtex4}
\usepackage{graphicx}% Include figure files
\usepackage{dcolumn,color}% Align table columns on decimal point
\usepackage{bm}% bold math

\begin{document}

%\preprint{MPF_DRAFTv3.0/29.09.2002}

\boldmath
\title{Modification of the $\omega$-Meson Lifetime in Nuclear Matter}
\unboldmath

\author{
  M.~Kotulla~$^1$,
  D.~Trnka~$^1$,
  P.~M\"uhlich~$^{1a}$,
  G.~Anton~$^2$,
  J.~C.~S.~Bacelar~$^3$,
  O.~Bartholomy~$^4$,
  D.~Bayadilov~$^{4,8}$,
  Y.A.~Beloglazov~$^8$,
  R.~Bogend\"orfer~$^2$,
  R.~Castelijns~$^3$,
  V.~Crede~$^{4,^\ast}$,
  H.~Dutz~$^5$,
  A.~Ehmanns~$^4$,
  D.~Elsner~$^5$,
  R.~Ewald~$^5$,
  I.~Fabry~$^4$,
  M.~Fuchs~$^4$,
  K.~Essig~$^4$,
  Ch.~Funke~$^4$,
  R.~Gothe~$^{5,^\diamond}$,
  R.~Gregor~$^1$,
  A.~B.~Gridnev~$^8$,
  E.~Gutz~$^4$,
  S.~H\"offgen~$^5$,
  P.~Hoffmeister~$^4$,
  I.~Horn~$^4$,
  J.~H\"ossl~$^2$,
  I.~Jaegle~$^7$,
  J.~Junkersfeld~$^4$,
  H.~Kalinowsky~$^4$,
  Frank~Klein~$^5$,
  Fritz~Klein~$^5$,
  E.~Klempt~$^2$,
  M.~Konrad~$^5$,
  B.~Kopf~$^{6,9}$,
  B.~Krusche~$^7$,
  J.~Langheinrich~$^{5,^\diamond}$,
  H.~L\"ohner~$^3$,
  I.V.~Lopatin~$^8$,
  J.~Lotz~$^4$,
  S.~Lugert~$^1$,
  D.~Menze~$^5$,
  J.~G.~Messchendorp~$^3$,
  T.~Mertens~$^7$,
  V.~Metag~$^1$,
  U.~Mosel~$^{1a}$,
  M.~Nanova~$^1$,
  R.~Novotny~$^1$,
  M.~Ostrick~$^5$,
  L.~M.~Pant~$^{1,^\dagger}$,
  H.~van Pee~$^1$,
  M.~Pfeiffer~$^1$,
  A.~Roy~$^{1,^\ddagger}$,
  A.~Radkov~$^8$,
  S.~Schadmand~$^{1,^\star}$,
  Ch.~Schmidt~$^4$,
  H.~Schmieden~$^5$,
  B.~Schoch~$^5$,
  S.~Shende~$^3$,
  G.~Suft~$^2$,
  V.~V.~Sumachev~$^8$,
  T.~Szczepanek~$^4$,
  A.~S\"ule~$^5$,
  U.~Thoma~$^{1,4}$,
  R.~Varma~$^{1,^\ddagger}$,
  D.~Walther~$^5$,
  Ch.~Weinheimer~$^{4,^+}$,
  Ch.~Wendel~$^4$\\
(The CBELSA/TAPS Collaboration)
}
\affiliation{
  $^1$II. Physikalisches Institut, Universit\"at Giessen, Germany\\
  $^{1a}$Institut f\"ur Theoretische Physik, Universit\"at Giessen, Germany\\
  $^2$Physikalisches Institut, Universit\"at Erlangen, Germany\\
  $^3$KVI, Groningen, The Netherlands\\
  $^4$\mbox{Helmholtz-Institut f\"ur Strahlen- u. Kernphysik, Universit\"at Bonn, Germany}\\
  $^5$Physikalisches Institut, Universit\"at Bonn, Germany\\
  $^6$\mbox{Institut f\"ur Kern- und Teilchenphysik, TU Dresden, Germany}\\
  $^7$Physikalisches Institut, Universit\"at Basel, Switzerland\\
  $^{8}$Petersburg Nuclear Physics Institute, Gatchina, Russia\\
  $^9$Physikalisches Institut, Universit\"at Bochum, Germany\\
  $^{\diamond}$ now at University of South Carolina, Columbia, USA\\
  $^\ast$ now at Florida State University, USA\\
  $^\star$\mbox{now at Institut f\"ur Kernphysik, FZ J\"ulich, Germany}\\
  $^+$\mbox{now at Institut f\"ur Kernphysik, Universit\"at M\"unster, Germany}\\
  $^\dagger$ \mbox{on leave from Nuclear Physics Division, BARC, Mumbai, India} \\
  $^\ddagger$ \mbox{on leave from Department of Physics, I.I.T. Powai, Mumbai, India}\\
  }%
\date{\today}% It is always \today, today,

\begin{abstract}
The photo production of $\omega$ mesons on the nuclei C, Ca, Nb and Pb has been
measured using the Crystal Barrel/TAPS detector
at the ELSA tagged photon facility in Bonn. The dependence of the $\omega$
meson cross section on the nuclear mass number has been compared with three different types of models, a
Glauber analysis, a BUU analysis of the Giessen theory group
and a calculation by the Valencia theory group.
In all three cases, the inelastic $\omega$ width is found to be
$130-150~\rm{MeV/c^2}$ at normal nuclear matter density for an average
3-momentum of 1.1~GeV/c. 
In the restframe of the $\omega$ meson, this inelastic $\omega$ width
corresponds to a reduction of the $\omega$
lifetime by a factor $\approx 30$.
For the first time, the momentum dependent $\omega$N cross section has been
extracted from the experiment and is in the range of $70$ mb.
\end{abstract}

\boldmath
\maketitle
\unboldmath
The investigation and understanding of in-medium properties of hadrons has
advanced to one of the most attractive research topics in hadron physics.
Several theoretical studies (e.~g. \cite{meis,brown,hats,leup,weise}) have led
to the expectation of a partial restoration of chiral symmetry at high
temperatures or with increasing nuclear densities.
A consistent picture of corresponding in-medium changes of hadrons 
has, however, not yet emerged.
These investigations have
stimulated a series of 
measurements to study the effect of surrounding strongly interacting matter
on the mass and width of hadrons.
%\par
Recently, several experiments have focused on studying the in-medium
properties of vector mesons. For
$\rho$ mesons a broadening but no mass shift has been reported in cold \cite{jlab} and heated \cite{na60}
nuclear matter. In contrast, the authors in \cite{kek} report a mass shift but no broadening of the $\rho$ meson
in nuclear matter. Also for $\phi$ mesons with low momenta a lowering of the in-medium mass has been reported
\cite{kek1}
as well as for the $\omega$ meson \cite{trnka}. It is important to note that
the analysis of the $\omega$ data \cite{trnka} was not sensitive to the
in-medium decay width 
due to the detector resolution and uncertainties in the separation of in- and
out-of-medium decay contributions.\\
%\par
 This paper describes an access to the in-medium
width of the $\omega$ meson via the measurement of the transparency ratio. This method has been
motivated in earlier works on the $\phi$ meson \cite{cab,mue}
as well as more recently on the $\omega$ meson \cite{kas,mue1}.
Experimentally, the LEPS collaboration at Spring8 extracted the
 transparency ratio for the photo production of $\phi$ mesons \cite{lep} and reported an
unexpectedly large in-medium inelastic cross section of $\phi$ mesons at normal nuclear matter density.
\par
\begin{figure}
%  \vspace*{-1.2cm}
  \includegraphics[width=0.9\columnwidth]{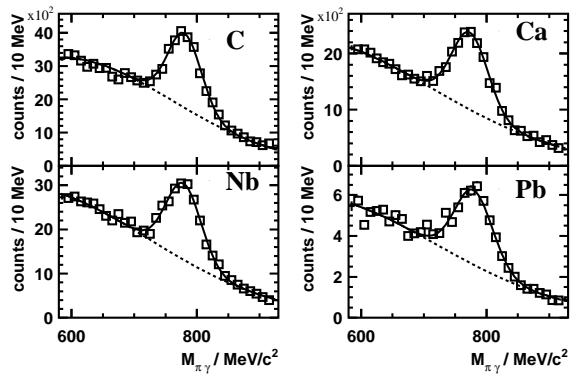}\hspace*{-0.1cm}
%   \vspace*{-0.6cm}
  \caption[signal]{Invariant mass distributions
 of a $\pi^0 \gamma$ pair for the four different targets integrated over
all momentum bins. The $\omega$ signal
 has been
 extracted by fitting a Gaussian function to the signal on top of a background function.
\vspace*{-0.4cm}
\label{signal}}
\end{figure}
\par
\par
The present experiment was performed at the
{\bf EL}ectron {\bf S}tretcher {\bf A}ccelerator
(ELSA) in Bonn, using a 2.8 GeV electron beam. The photon beam was produced
via bremsstrahlung.
A magnetic spectrometer (tagger)
was used to determine the photon beam energies within the tagged photon range of 0.64 to 2.53 GeV.
The C, Ca, Nb, and Pb targets had thicknesses of 20 mm, 10 mm, 1~mm and 0.64~mm, respectively, and
30 mm in diameter.
The targets were mounted in the center of the Crystal Barrel detector (CB), a photon
calorimeter consisting of 1290 CsI(Tl) crystals ($\sim$~16~
radiation lengths $X_0$) with an angular coverage of $30^\circ$ up to $168^\circ$ in the polar angle
and a complete azimuthal angular coverage. Inside the CB, covering its full acceptance,
a three-layer scintillating fiber detector
(513 fibers of 2~mm diameter) was installed for charged particle detection.
Reaction products emitted in forward direction
were detected in the TAPS detector. TAPS consisted of
528 hexagonally shaped $\rm{BaF_2}$ detectors
($\sim$~12~$X_0$)
covering polar angles between $4^\circ$ and
$30^\circ$ and the complete $2\pi$ azimuthal angle.
In front of each $\rm{BaF_2}$ module a 5 mm thick plastic scintillator
was mounted for the registration of charged particles.
The resulting geometrical solid angle coverage of the combined system was $ 99 \%$ of $4\pi$.
The $\rm{BaF_2}$ crystals were read out by
photo multipliers providing a fast trigger, the CsI(Tl) crystals via photo diodes.
For further details see \cite{cb,taps,taps1}.
\par
The experimental observable for extracting the in-medium width is the {\it transparency ratio}, defined as:
\begin{eqnarray}\label{ta}
T=\frac{\sigma_{\gamma A\rightarrow VX}}{A\;\sigma_{\gamma
N\rightarrow V X}}~,
\end{eqnarray}
i.~e. the ratio of the inclusive nuclear $\omega$ photo production cross section
divided by $A$ times the same quantity on a free nucleon.
$T$ describes the loss of
flux of $\omega$ mesons in nuclei and is related to the
absorptive part of the $\omega$ nucleus
potential and thus to the $\omega$ inelastic width in the nuclear medium.
To avoid systematic uncertainties when comparing to theoretical models, e.~g.
due to the unknown $\omega$ production cross section on the neutron or secondary production
processes, the transparency ratio has been normalized to the Carbon data, i.~e.
\begin{eqnarray}\label{tac}
T_A=\frac{12 \cdot \sigma_{\gamma A\rightarrow VX}}{A\;\sigma_{\gamma
C_{12}\rightarrow V X}}~.
\label{eq:trans}
\end{eqnarray}
The result is thereby normalized to a light target with equal numbers of
protons and neutrons. 
\begin{figure}
  \includegraphics[width=0.99\columnwidth]{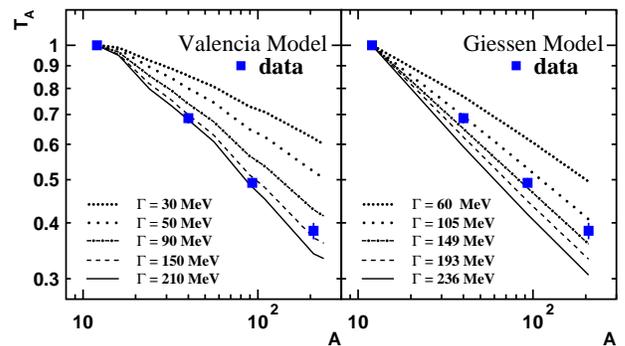}%\hspace*{-0.5cm}
  \caption[womomentum]{
(color online) Experimentally determined transparency ratio according to
Eq.~\ref{eq:trans} in comparison with a theoretical Monte Carlo simulation
\cite{kas2,kas3} (left) and a BUU calculation \cite{muehl2,muehl3} (right)
varying the width at 1.1~GeV/c momentum,
respectively. The width is given in the nuclear restframe.   
Only statistical errors are shown.
\label{wo_mom}}
\vspace*{0.047cm}
\end{figure}
\par
The $\omega$ meson has been reconstructed via the mode
$\omega \rightarrow \pi^0\gamma$ with the $\pi^0$ further decaying into 2 photons, thus requiring three
neutral hits in the CB/TAPS detector systems. An identical analysis code has been used for all four
nuclear targets. In the analysis incident photon beam energies in the range of 1.2 GeV to 2.2 GeV have
been allowed for.
The photon flux has been determined by counting the scattered electrons in the
tagging system and by correcting for the tagging efficiency. To
reduce background a cut on the energy of the decay photon not belonging to the $\pi^0$ has been set,
$E_{\gamma3} > 200$~MeV. Events distorted by final state interactions of the $\pi^0$ meson
have been removed by applying a cut on the
kinetic energy of the decay pions, $T_{\pi^0} > 150$~MeV (see e.g. \cite{messch,muehl2}).
The resulting $\pi^0\gamma$ invariant mass distributions are shown in
Fig. \ref{signal} for all four nuclear targets. The $\omega$ signal
 has been  extracted by fitting a Gaussian function to the signal on top of a background function.
Since almost all $\omega$ mesons decay outside the
nuclear target at average momenta of 1.1~GeV/c \cite{trnka}, neither a direct mass shift nor a
 broadening of the width can be observed in the $\pi^0 \gamma$ invariant mass distributions 
(Fig.~\ref{signal}). The statistical error of the fitting procedure was
taken to be $\sqrt{(S+2B)}$ using the extracted signal $S$ and background
$B$ events.
The systematic errors of the extracted transparency
ratios (Eq.~\ref{ta}) include the uncertainties of the target length, the
fraction of coherent $\omega$ production and the $\gamma$ conversion in the
target, respectively, for each target.
Contributions from coherent $\omega$ production are negligible, since the
transparency ratios differ by less than $2\%$ when comparing fully
inclusive $\omega$ production (including coherent contributions) to quasi-free
production with detection of a recoil proton. 
The quadratic addition of these
individual errors yields a
systematic uncertainty of 3.5\%.
\par
\par
Knowledge of the photon flux, efficiency, and the number of measured $\omega$ mesons is
sufficient to deduce the partial $\omega \rightarrow \pi^0\gamma$ decay cross section.
The dependence of the cross section on the nuclear mass number $A$ can be
parameterized as:%   $\sigma(p_\omega,A)~\propto A^{\alpha(p_\omega)}$.
\begin{equation}
  \sigma(p_\omega,A)~\propto A^{\alpha(p_\omega)}
  \label{eq:alpha}
\end{equation}
The measured cross sections are in good agreement with an $A^\alpha$~scaling law.
The fit to the data yields $\alpha$ values between $0.54 - 0.74$ from
lower to higher 3-momenta $p_\omega$.  This result is in
agreement with the data
obtained by the KEK collaboration \cite{kek2}
%($\alpha~=0.710 \pm 0.021 (stat) \pm 0.037 (syst)$)
and indicates a strong absorption of $\omega$ mesons in the nuclear medium.
\par
In a next step, the transparency ratio as defined in Eq.~\ref{eq:trans} has been extracted and
is shown in Fig.~\ref{wo_mom} as full (blue) circles.
The data in Fig. \ref{wo_mom} are compared to two theoretical models, a
Monte Carlo type analysis by the Valencia group \cite{kas3}
(left panel) and a BUU transport code calculation by the Giessen group
\cite{muehl3} (right panel).
\begin{figure}
  \includegraphics[width=0.7\columnwidth]{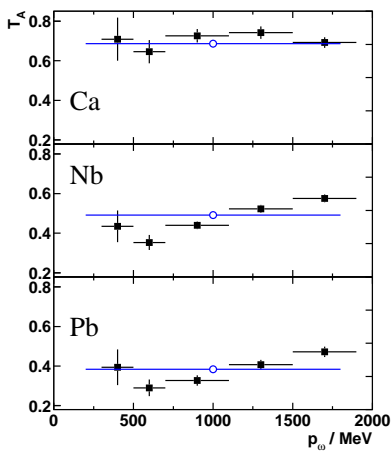}\hspace*{-0.1cm}
  \caption[momentum]{(color online) Momentum dependent transparency
    ratio normalized to C for the three different targets Ca, Nb and Pb as full squares. The open circles show
the values when integrating over all momenta and correspond to the data points shown in figure \ref{wo_mom}.
Only statistical error bars are shown.
    More details are given in the text.
\vspace*{-0.4cm}
\label{momdep}}
\end{figure}
The collision width included in the Giessen BUU model is evaluated
consistently from the $\omega$ collision rates using the low-density
approximation and, thus, is momentum dependent, while the Valencia group
uses a momentum-independent width. The values given in Fig. \ref{wo_mom}
correspond to $\omega$ mesons with momenta of 1.1~GeV/c which is the
average momentum of the $\omega$ mesons in the data sample. A comparison
of the data to the model predictions yields an in-medium $\omega$
width in the nuclear restframe of $\rm{130 - 150~MeV/c^2}$.
In addition, we have investigated the momentum dependence of the transparency
ratio to extract
the momentum dependence of the $\omega$ meson inelastic collision width.
These data normalized to C for the three different
targets Ca, Nb and Pb are shown in
Fig. \ref{momdep} as full squares. The open circles show
the values when integrating over all momenta and correspond to the data points shown in Fig. \ref{wo_mom}.
For the heavier targets the data indicate a slight
decrease of the transparency ratio with decreasing $\omega$ 3-momenta.
\par
We have used the Glauber model in the high energy eikonal approximation to
analyze these data points and to extract the momentum dependence of the
$\omega$ meson inelastic width. The Glauber model \cite{Glauber:1970jm}
was first applied to photoproduction experiments by Margolis
\cite{Margolis:1968} to extract the inelastic $\rho$N cross section both
from coherent and incoherent $\rho$ photoproduction off nuclei. A very
detailed description and application of the Glauber model to
photoproduction reactions can be found in \cite{Bauer:1977iq}.
 More recently, the Glauber model has been applied to
study the effects of color transparency \cite{Kopeliovich:1993pw} and
nuclear shadowing \cite{Falter:2000yi} in high energy photonuclear
reactions. In the Glauber eikonal approximation, neglecting Fermi motion
and Pauli blocking, the total incoherent cross section for the
photoproduction of a single vector meson is given by
\begin{eqnarray}\label{eq_sigma}
\lefteqn{\frac{d\sigma_{\gamma A\to VX}}{dp}}  \\  &=&
\frac{d\sigma_{\gamma N\to VN}}{dp} \, \int d^3r\,\rho_N({\bf r}^{\,})
\exp\left(-\int\limits_{z}^{\infty}\frac{dP_{\rm
abs}(z')}{dl}dz'\right)\nonumber
\end{eqnarray}
where $dP_{\rm abs}/dl$ denotes the absorption probability per unit length
of the vector meson in nuclear matter. The absorption probability per unit
length is related to the inelastic width in the rest frame of nuclear
matter via
\begin{eqnarray}
\frac{dP_{\rm abs}}{dl}=\frac{dP_{\rm abs}}{dt}\frac{dt}{dl}=\frac{\Gamma_{\rm inel}}{v}=\frac{E_V}{p}\Gamma_{\rm inel}
\end{eqnarray}
with the vector meson energy $E_V$ at 3-momentum $p$ and velocity $v$.
Here, the inelastic width depends on density and on the velocity of the
$\omega$ meson. Using the low-density approximation we can separate the
density- and momentum-dependence
\begin{eqnarray}
\Gamma_{\rm inel}(\rho_N,p)=\Gamma_{0}(p)\,\frac{\rho_N}{\rho_0}
\label{eq:width}
\end{eqnarray}
with $\rho_0=0.16~{\rm fm}^{-3}$.
Carrying out some of the integrals in Eq.~(\ref{eq_sigma}), the nuclear photoproduction cross section can be written as
%\begin{eqnarray}
%\frac{d\sigma_{\gamma A\to \omega X}}{dp}=A_{\rm eff}(p)\frac{d\sigma_{\gamma N\to \omega X}}{dp}
%\end{eqnarray}
%with
%\begin{eqnarray}\label{eq_aeff}
%A_{\rm eff}(p)=2\pi\rho_0\frac{p}{E_V\Gamma_{0}(p)}\times
% \;\;\;\;\;\;\;\;\;\;\;\;\;\;\;\;\;\;\;\;\;\;\;\;\;\;\;\;\;\;\;\;\;\;\;\;\;\;\; \nonumber\\
%\int\limits_0^{\infty} b\,db \left[1-\exp\left(-\frac{E_V}{p}\int\limits_{-\infty}^{+\infty}dz'\,
%\Gamma_{\rm inel}(\{{\bf b},z'\},p)\right)\right]
%\end{eqnarray}
\begin{eqnarray}\label{eq_aeff}
\lefteqn{\frac{d\sigma_{\gamma A\to \omega X}}{dp} =\frac{d\sigma_{\gamma
N\to \omega X}}{dp} \times \rho_0\frac{p}{E_V\Gamma_{0}(p)}}
\\
&\times& 2\pi\int\limits_0^{\infty} b\,db
\left[1-\exp\left(-\frac{E_V}{p}\int\limits_{-\infty}^{+\infty}dz'\,
\Gamma_{\rm inel}(\{{\bf b},z'\},p)\right)\right] \nonumber
\end{eqnarray}
Note, that the width $\Gamma_{\rm inel}$ in the above equation
relates to the rest frame of nuclear matter. Equation (\ref{eq_aeff}) has
been used in order to fit the measured transparency ratios shown in
Fig.~\ref{momdep} for fixed values of the $\omega$ meson 3-momentum by
varying the $\omega$ meson inelastic width. In this way the widths
$\Gamma_0(p)$ and their errors were determined separately for each target.
 \par
\begin{figure}
  \hspace*{-0.2cm}
  \includegraphics[width=0.95\columnwidth]{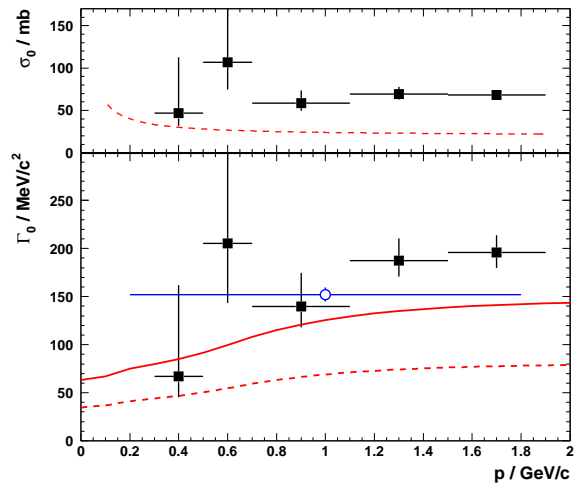}\hspace*{-0.1cm}
   \vspace*{-0.06cm}
  \caption[signal]{(color online). Upper part: The inelastic $\omega$N
    cross section 
    extracted from the Glauber analysis (data) in comparison to the
    inelastic cross 
    section used in the BUU simulation \cite{muehl2,muehl3}. Lower part:
    Width of the $\omega$ 
    meson in the nuclear
    medium in the nuclear restframe as a function of the $\omega$ momentum in a
    Glauber analysis
    (squares), from the Giessen BUU model with the inelastic
    cross section from the upper figure (red dashed line) and after fit to the
    data of Fig.~\ref{wo_mom} with BUU (red line), 
    and the Valencia Monte
    Carlo simulation (blue circle), respectively.
    Only statistical errors are shown.
\vspace*{-0.4cm}
\label{glauber}}
\end{figure}

The values for the individual targets were combined into an error weighted
average and are shown in in the lower part of Fig.~\ref{glauber}.
For the different momentum bins an increase of the $\omega$ width up to
momenta of about 1~GeV/c is observed. The extracted momentum behavior from
the Glauber model is in reasonable agreement with the BUU
parameterization. In contrast, the Monte Carlo simulation by the Valencia
group explicitly assumes no momentum dependence. For a comparison of the
$\omega$ in-medium width with the $\omega$ width in vacuum, the values
shown in Fig.~\ref{wo_mom},~\ref{glauber} have to be transformed to the
$\omega$ restframe.
The observed width of 130-150~MeV/c$^2$ at 1.1~GeV/c momentum corresponds to a
width in the restframe of the $\omega$ meson of $225-260$~MeV/c$^2$ .
The upper part of Fig.~\ref{glauber} contains the first determination of the
inelastic $\omega N$ cross 
 section extracted by using the classical low-density relation
\vspace{-5mm}
\begin{equation}
%\Gamma_0(p) = \rho_0 \frac{p}{E_V} \sigma_0(p)
\sigma_0(p) = \frac{\Gamma_0(p)}{\rho_0} \frac{E_\omega}{p}
\label{eq:low}
\end{equation}
%\vspace{-2mm}
Similar to
the study of the $\phi$ meson properties in the nuclear medium \cite{lep}, the
cross section at large momenta exceeds by a factor $\approx 3$ the
inelastic $\omega N$ cross section \cite{Lyk} used as input in the BUU
calculations \cite{muehl2,muehl3}.  

\par
In summary, we have deduced the in-medium width of the $\omega$ meson from photo production experiments
using the Crystal Barrel/TAPS detector systems at the ELSA accelerator facility in Bonn. It is found that the
$\omega$ eigenlifetime decreases by a factor $\approx 30$ for normal nuclear
matter density 
compared to the vacuum value. Furthermore, the
momentum dependence of the transparency ratio and of the extracted in-medium
$\omega$ width has been determined for the first time. Deviations of the
experimental data 
from the momentum dependence of the $\omega$ in-medium width implemented as
input in
the BUU code indicate 
room for improvement in the parameterization of the $\omega$N cross
section or a limited applicability of the low density theorem (Eq.~\ref{eq:low}).
\par
We gratefully acknowledge stimulating discussions with E. Oset
and M.~Kaskulov. We thank the accelerator group of ELSA as well as
the technicians and scientists of the HISKP in Bonn, the PI in Bonn and the
II. Physikalisches Institut in Giessen.
This work was supported by the Deutsche Forschungsgemeinschaft, SFB/TR-16, and the Schweizerischer Nationalfond.
%\bibliography{./mybibliography}% Produces the bibliography via BibTeX.

\end{document}